\documentclass[12pt]{article}
\usepackage{graphicx}
\usepackage{caption}
\usepackage{subcaption}
\usepackage{epstopdf}
\usepackage{epsfig}
\usepackage{amsmath}
\begin{document}

\title{Partial quark-lepton universality \\and neutrino CP violation }
\author{Jiajun Liao$^{1,2}$, D. Marfatia$^{1}$, and K. Whisnant$^2$\\
\\
\small\it $^1$Department of Physics and Astronomy, University of Hawaii at Manoa, Honolulu, HI 96822, USA\\
\small\it $^2$Department of Physics and Astronomy, Iowa State University, Ames, IA 50011, USA
}
\date{}
\maketitle

\begin{abstract}
We study a model with partial quark-lepton universality that can naturally arise in grand unified theories. We find that constraints on the model can be reduced to a single condition on the Dirac CP phase $\delta$ in the neutrino sector. Using our current knowledge of the CKM and PMNS mixing matrices, we predict $-32.4^\circ\leq\delta\leq 32.0^\circ$ at $2\sigma$.
\end{abstract}
\newpage
\section{Introduction}
Our understanding of neutrinos has progressed steadily in the last two decades. After the observation of nonzero $\theta_{13}$ by the Daya
Bay~\cite{An:2012eh,daya2}, RENO~\cite{Ahn:2012nd}, and Double Chooz~\cite{DC} experiments, we now know the three mixing angles $\theta_{12}$, $\theta_{23}$, and $\theta_{13}$, and the two mass squared differences to good precision. For the normal hierarchy, current $2\sigma$ ranges of the three mixing angles from a global three-neutrino oscillation analysis are~\cite{Capozzi:2013csa}

\begin{align}
\theta_{12}=33.7^{+2.1}_{-2.1}(^\circ)\,,\ \ \ \ \theta_{23}=41.4^{+6.6}_{-2.6}(^\circ)\,, \ \ \ \  \theta_{13}=8.80^{+0.73}_{-0.77}(^\circ)\,.
\end{align}

The focus of next generation neutrino oscillation experiments is shifted to the Dirac CP phase $\delta$ and the neutrino mass hierarchy.  Predictions of the many theoretical models designed to explain the observed mixing patterns await verification. Among these models, quark-lepton universality (QLU)~\cite{Joshipura:2005cm} is well-motivated. It is based on simple relations in Grand Unified Theories (GUT) and connects the mixing matrices of quarks and leptons. Exact quark-lepton universality leads to a symmetric PMNS mixing matrix. However, using the current $3\sigma$ ranges of the oscillation parameters~\cite{Capozzi:2013csa}, we find the moduli of the neutrino mixing matrix elements are
 \begin{equation}
 |V_{PMNS}| = \begin{pmatrix}
   0.789 - 0.853 & 0.501 - 0.594 & 0.133 - 0.172 \\
   0.195 - 0.556 & 0.410 - 0.733 & 0.602 - 0.784 \\
   0.196 - 0.557 & 0.411 - 0.733 & 0.602 - 0.784
   \end{pmatrix}\,.
\end{equation}
We see that the exactly symmetric PMNS mixing matrix is disfavored by current data. This aspect of the PMNS matrix with $V_{PMNS}=V_{PMNS}^T$ or $V_{PMNS}=V_{PMNS}^\dagger$ has been studied in Refs.~\cite{Hochmuth:2006xa,BenTov:2012wa,Guo:2012xb}.

In this Letter, we discuss partial quark-lepton universality~\cite{Joshipura:2005cm}, which does not require the unitary matrices that diagonalize the upper and lower components of the weak doublets to be the same.  We find that partial QLU fits the current data very well and we can make a prediction for the unknown Dirac CP phase.


In Section 2, we review partial quark-lepton universality and discuss renormalization group effects on the model. In Section 3, we discuss the phenomenological results of this model and predict the Dirac CP phase. We conclude in Section~4. 

\section{Partial quark-lepton universality}
Partial quark-lepton universality can be derived from some simple relations in grand unified theories~\cite{Joshipura:2005cm}. We start with the SU(5) relation,
\begin{align}
M_l=M_d^T\,,
\label{eq:lower}
\end{align}
obtainable in lopsided models~\cite{barr}, and 
\begin{align}
M_u=M_u^T\,,
\label{eq:upper}
\end{align}
where $M_l$, $M_u$, and $M_d$ are the mass matrices of the charged-leptons, up-type quarks and down-type quarks, respectively. If we assume $M_d$ is Hermitian, which can be achieved
by imposing left-right symmetry~\cite{Joshipura:2005cm},{\footnote{Implementing an Hermitian $M_d$ in a GUT is difficult because SU(5) does not incorporate left-right symmetry, and in SO(10), the mass matrices arising from the couplings of fermions to Higgs fields in the 10 and 126 representations are complex symmetric (and not Hermitian), while those arising from couplings to 120 are complex antisymmetric.}  then from Eq.~(\ref{eq:lower}) we find that both the down-type quarks and charged-leptons can be diagonalized by a unitary matrix $V$, 
\begin{align}
V^\dagger M_d V=D_d\,,\ \ \ \ V^T M_l V^*=D_l\,.
\label{eq:dl}
\end{align}

Also, from Eq.~(\ref{eq:upper}) we know that the up-type quarks can be diagonalized by a unitary matrix $V'$,
\begin{align}
V'^\dagger M_u V'^*=D_u\,.
\label{eq:u}
\end{align}
If the Dirac neutrino matrix $M_{\nu D}$ and the right-handed Majorana neutrino mass matrix $M_R$ are also diagonalized by $V'$ (as in some SO(10) models~\cite{Joshipura:2005cm}),
\begin{align}
V'^\dagger M_{\nu D} V'^*=D_{\nu D}\,,\ \ \ \  V'^\dagger M_R V'^*=D_R\,,
\label{eq:nud}
\end{align}
 then below the seesaw scale, the light neutrino mass matrix, $M_\nu=-M_{\nu D}M_R^{-1}M_{\nu D}^T$, is diagonalized by $V'$ as well,
\begin{align}
V'^\dagger M_{\nu} V'^*=D_{\nu}\,.
\label{eq:nu}
\end{align}
From Eqs.~({\ref{eq:dl}}),~(\ref{eq:u}) and~(\ref{eq:nu}), we can find that the observable mixing matrices are related by
\begin{align}
V_{CKM}=V'^\dagger V\,,
\label{eq:CKM}
\end{align}
and
\begin{align}
V_{PMNS}=V^TV\,'.
\label{eq:PMNS}
\end{align}


Note that for exact quark-lepton universality, we must have $V'=V$, which indicates that $V_{CKM}=I$ and the $V_{PMNS}$ mixing matrix is symmetric. This is disfavored by current data. In the next section, we show that partial quark-lepton universality is still allowed by current data. A caveat to partial QLU is that small perturbations to the leading order relations of Eqs.~(\ref{eq:CKM}) and~(\ref{eq:PMNS}) are needed to reproduce the measured fermion masses. In Ref.~\cite{Joshipura:2005cm} it was shown that with a specific form for the perturbations, the measured fermion masses can be obtained while keeping the mixing matrices unchanged. Consequently, we focus on the connection between the mixing matrices of quarks and leptons.{\footnote{
An example in which Eqs.~({\ref{eq:lower}}),~(\ref{eq:upper}),~(\ref{eq:nud}) and the Hermiticity of $M_d$ naturally arise, is an SO(10) scheme with the superpotential terms~\cite{Joshipura:2005cm},
\begin{align}
W_d=\frac{f_{ij}}{M}(16_i^T B \varGamma_\mu H)(H'^T B\varGamma_\mu 16_j)+\frac{f'_{ij}}{M}(16_i^T B \varGamma_\mu H')(H^T B\varGamma_\mu 16_j)
\end{align}
and 
\begin{align}
W_u=g_{ij}(16_i^TB\varGamma_{\mu\nu\lambda\sigma\rho}16_j)\Phi^{\mu\nu\lambda\sigma\rho}\,,
\end{align}
where $H$, $H'$ are 16-plet Higgses, $\Phi$ is a $\overline{126}$-plet Higgs, B is a charge conjugation matrix in SO(10), $i$ and $j$ are generation indices, and $\mu,\nu,\lambda,\sigma,\rho$ are SO(10) indices. The Lorentz indices and the standard charge conjugation matrix are suppressed. $H$ and $H'$ contain neutral fields with the quantum numbers of $\nu$ and $\nu^c$, so that the vacuum expectation value for $\nu^c$ breaks SO(10) while SU(5) is preserved. We take the $\overline{126}$ contribution to $H_d$ to be zero or sub-dominant compared to $H$ and $H'$, so $M_d$ is only generated from $W_d$. By imposing an additional symmetry, $16\rightarrow 16^*,H\rightarrow H'^*$, which leads to $f_{ij}\rightarrow f'^*_{ij}$,  an Hermitian $M_d$ can be obtained.}}

Current data that determine the CKM and PMNS mixing matrices are measured at low energies, while the quark-lepton universality relations are realized at the grand unification scale. In order to use current data to analyze the model, we must consider renormalization group (RG) effects. For the CKM matrix, the RG effects are very small, i.e., the next order relative corrections to the CKM matrix are of the order $\lambda^5$~\cite{Kielanowski:2000cn,lindner}, where $\lambda=0.225$. The RG effects in the neutrino sector are strongly dependent on the mass spectrum of the light neutrinos. For the inverted and quasi-degenerate mass hierarchies, the effects can be large~\cite{Casas:1999tg,balaji,haba,chanowski}. However, with quark-lepton universality it is more natural to assume that the light neutrinos are very hierarchical with the normal mass spectrum. In this case, RG effects on the three angles are very small~\cite{Antusch:2005gp,schmidt}, e.g., $\delta\theta_{23}\sim 0.6^\circ$, $\delta\theta_{13}\sim 0.2^\circ$ and  $\delta\theta_{12}\sim 0.8^\circ$ in the MSSM with $\tan\beta=20$ if the lightest neutrino mass is 0.01~eV. Since current uncertainties in the three angles are larger than the RG effects, we neglect the RG effects in our analysis.
 
\section{Phenomenology}
In this section, we introduce a simple approach based on the properties of unitary matrices to reduce the constraints on the model to a single condition, which allows us to easily constrain the Dirac CP phase.

Partial QLU predicts the two observable mixing matrices to have the form of Eqs.~(\ref{eq:CKM}) and~(\ref{eq:PMNS}), which can be rewritten as
\begin{align}
V_{PMNS}V_{CKM}=V^TV\,,
\label{eq:pc}
\end{align}
and
\begin{align}
V_{CKM}^*V_{PMNS}=V'^TV'\,.
\label{eq:cp}
\end{align}
Hence, in order for the model to work, both $V_{PMNS}V_{CKM}$ and $V_{CKM}^*V_{PMNS}$ should be symmetric. However, the two constraints are not independent. Since Eq.~(\ref{eq:CKM}) implies $V'=V V_{CKM}^\dagger$, Eq.~(\ref{eq:pc}) follows from Eq.~(\ref{eq:cp}). 

Solutions for $V$ and $V'$ will always exist because if $V_{CKM}^*V_{PMNS}$ is symmetric, then it can be diagonalized by a unitary matrix $U_s$, i.e., $U_s^TV_{CKM}^*V_{PMNS}U_s=D$, where $D$ is diagonal. This means that we can find the solution, $V'=\sqrt{D}U_s^\dagger$. Once $V'$ is known, the solution for $V$ can be obtained from Eq.~(\ref{eq:CKM}). Although solutions for $V$ and $V'$ exist, they are not unique. We can always insert a combination of a real rotation matrix $R^TR$ into the middle of the right-handed side of Eq.~(\ref{eq:pc}) or~(\ref{eq:cp}), and since $R^TR=I$, the equation will not change. This can also be seen from Eqs.~(\ref{eq:CKM}) and~(\ref{eq:PMNS}). For any real rotation matrix $R$, $RV$ and $RV'$ are also unitary, hence if we let $V\rightarrow RV$ and $V'\rightarrow RV'$, the two observable mixing matrices will stay the same. 

Now if we define
\begin{align}
U=V_{CKM}^*V_{PMNS}\,,
\label{eq:U}
\end{align}
then the only constraint from the model is that $U$ is symmetric. Since both $V_{CKM}$ and $V_{PMNS}$ are unitary matrices, $U$ is also unitary. For a $3\times 3$ unitary matrix, it can be shown that $U$ being symmetric is equivalent to the moduli of $U$ being symmetric under phase redefinition~\cite{Branco:1990zq}. This constraint still imposes three conditions: $|U_{12}|=|U_{21}|$, $|U_{13}|=|U_{31}|$, and $|U_{23}|=|U_{32}|$. However, the conditions are not independent. Since $U$ is unitary, $|U_{11}|^2+|U_{12}|^2+|U_{13}|^2=|U_{11}|^2+|U_{21}|^2+|U_{31}|^2$, hence $|U_{12}|=|U_{21}|$ indicates $|U_{13}|=|U_{31}|$ and vice versa. Similarly, $|U_{23}|=|U_{32}|$ is equivalent to $|U_{13}|=|U_{31}|$. Therefore, there is only one independent condition that constrains the model. Here we choose it to be $|U_{13}|=|U_{31}|$.

The CKM matrix can be written in terms of the Wolfenstein parameters~\cite{Wolfenstein:1983yz} as,
\begin{align}
V_{CKM}=\begin{bmatrix}
   1-\lambda^2/2 & \lambda & A\lambda^3(\rho-i\eta) \\
   -\lambda & 1-\lambda^2/2 & A\lambda^2 \\
   A\lambda^3(1-\rho-i\eta) & -A\lambda^2 & 1
   \end{bmatrix}+\mathcal{O}(\lambda^4)\,,
\end{align}
and the PMNS matrix can be written in the standard form, which is 
\begin{align}
V_{PMNS}=\begin{bmatrix}
   c_{13}c_{12} & c_{13}s_{12} & s_{13}e^{-i\delta} \\
   -s_{12}c_{23}-c_{12}s_{23}s_{13}e^{i\delta} & c_{12}c_{23}-s_{12}s_{23}s_{13}e^{i\delta} & s_{23}c_{13} \\
   s_{12}s_{23}-c_{12}c_{23}s_{13}e^{i\delta} & -c_{12}s_{23}-s_{12}c_{23}s_{13}e^{i\delta} & c_{23}c_{13}
   \end{bmatrix},
\end{align}
where \(c_{ij}\), \(s_{ij}\) denotes \(\cos\theta_{ij}\) and \(\sin\theta_{ij}\) respectively, and Majorana phases are not included. From Eq.~(\ref{eq:U}), we see the condition $|U_{13}|=|U_{31}|$ becomes
\begin{align}
&|(1-\lambda^2/2)s_{13}e^{-i\delta}+\lambda s_{23}c_{13}+c_{23}c_{13}A\lambda^3(\rho+i\eta)| \nonumber \\
=&|A\lambda^3(1-\rho+i\eta)c_{13}c_{12}+A\lambda^2(s_{12}c_{23}+c_{12}s_{23}s_{13}e^{i\delta})+s_{12}s_{23}-c_{12}c_{23}s_{13}e^{i\delta}|\,.
\label{eq:cond}
\end{align}
Note that Eq.~(\ref{eq:cond}) cannot be satisfied when $\theta_{13}=0$.  Keeping in mind that $\sin\theta_{13}<\lambda$, the $\lambda^2 s_{13}^2$ and $\lambda^3 s_{13}$ terms can be neglected since they are of the same order of magnitude as the terms dropped in the Wolfenstein parametrization. Then we get a simple expression for the cosine of the Dirac CP phase:
\begin{align}
\cos\delta=\frac{s_{12}^2s_{23}^2+c_{12}^2c_{23}^2s_{13}^2-s_{13}^2-\lambda^2B}{2s_{23}c_{23}s_{12}c_{12}s_{13}+2\lambda s_{23}c_{13}s_{13}+2A\lambda^2s_{12}c_{12}(c_{23}^2-s_{23}^2)s_{13}}+\mathcal{O}(\lambda^4)\,,
\label{cosd}
\end{align}
where $B=s_{23}^2c_{13}^2-2As_{12}^2c_{23}s_{23}-2A\lambda(1-\rho)c_{12}s_{12}c_{13}s_{23}$. We see that for very small $\theta_{13}$ the numerator of the above equation is always larger than the denominator, so that there is no solution for $\delta$. 

Using the currently favored CKM~\cite{Bona:2005vz} and PMNS~\cite{Capozzi:2013csa} parameters with their respective uncertainties, and solving the condition $|U_{13}|=|U_{31}|$ numerically without any approximation,  we find the Dirac CP phase $\delta$ in the PMNS matrix to lie between $-32.4^\circ$ and $32.0^\circ$ at $2\sigma$. The asymmetry around 0 is due to the small CP violation in the CKM matrix, which does not enter the approximate result in Eq.~(\ref{cosd}).

We also find predictions for each mixing angle versus $\delta$ given the best-fit values and $2\sigma$ allowed regions of the other two mixing angles and the CKM parameters. The results are shown in Fig.~\ref{fg:delta}. With the constraints from the other two mixing angles and the CKM parameters, we find that $\theta_{23}<48.3^\circ$, $\theta_{12}<36.3^\circ$ and $\theta_{13}>7.64^\circ$ at $2\sigma$. 
The partial QLU model is perfectly consistent with current data, and rather large $\theta_{13}$ is strongly favored for the measured solar and atmospheric mixing angles. Note that the relevant neutrino mass squared differences are trivially accommodated.


A measurement of $\delta$ by future long baseline neutrino oscillation experiments will provide a stringent test of the viability of the partial quark-lepton universality model.

\section{Conclusion}
We studied partial quark-lepton universality, which can naturally arise in grand unified theories. Constraints on the model can be reduced to one simple condition, $|U_{13}|=|U_{31}|$. Dropping terms of order $\lambda^4$ from this condition, we find a simple expression for the Dirac CP phase $\delta$ in the neutrino sector. We also studied the allowed parameter regions of the model numerically. Our prediction that $\delta$ lies within the range $[-32.4^\circ, 32.0^\circ]$ at the $2\sigma$ level, will be tested by future long baseline neutrino experiments.
\vskip 0.1in
{\bf Acknowledgements}
\vskip 0.1in
This research was supported by the U.S. Department of Energy grant No. DE-SC0010504.

\begin{figure}
        \centering
        \begin{subfigure}[b]{0.59\textwidth}
                \includegraphics[width=\textwidth]{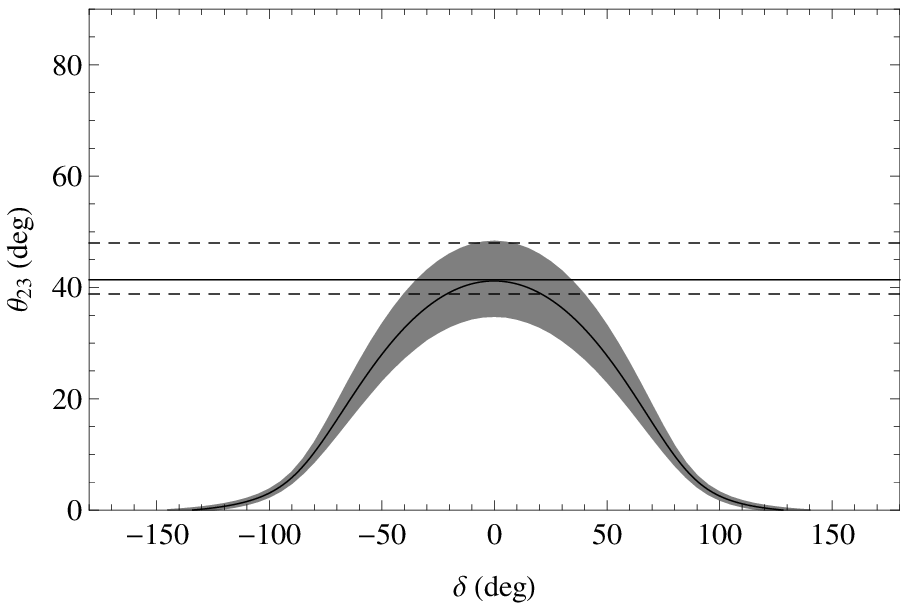}
        \end{subfigure}%
        \\
        \begin{subfigure}[b]{0.59\textwidth}
                \includegraphics[width=\textwidth]{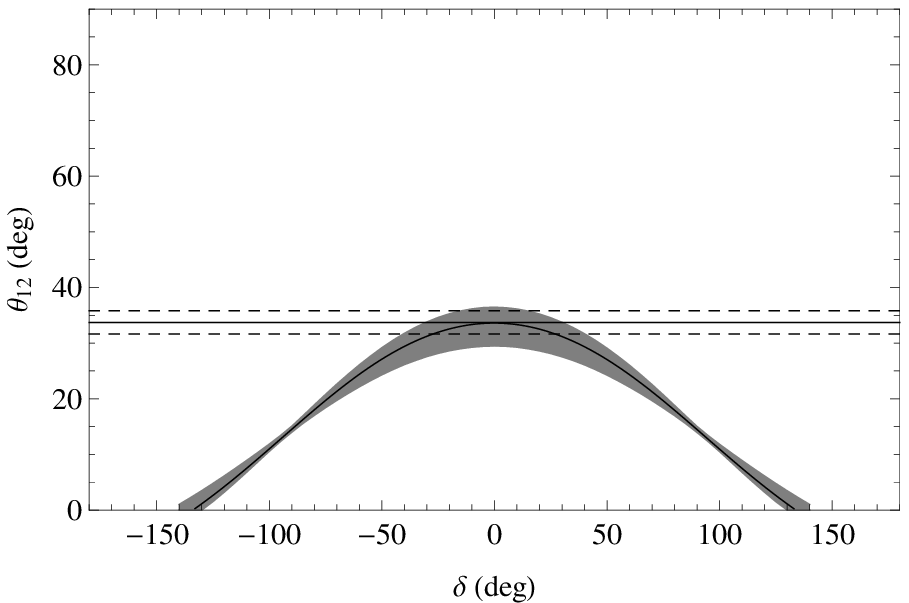}
        \end{subfigure}
        \\
        \begin{subfigure}[b]{0.59\textwidth}
                \includegraphics[width=\textwidth]{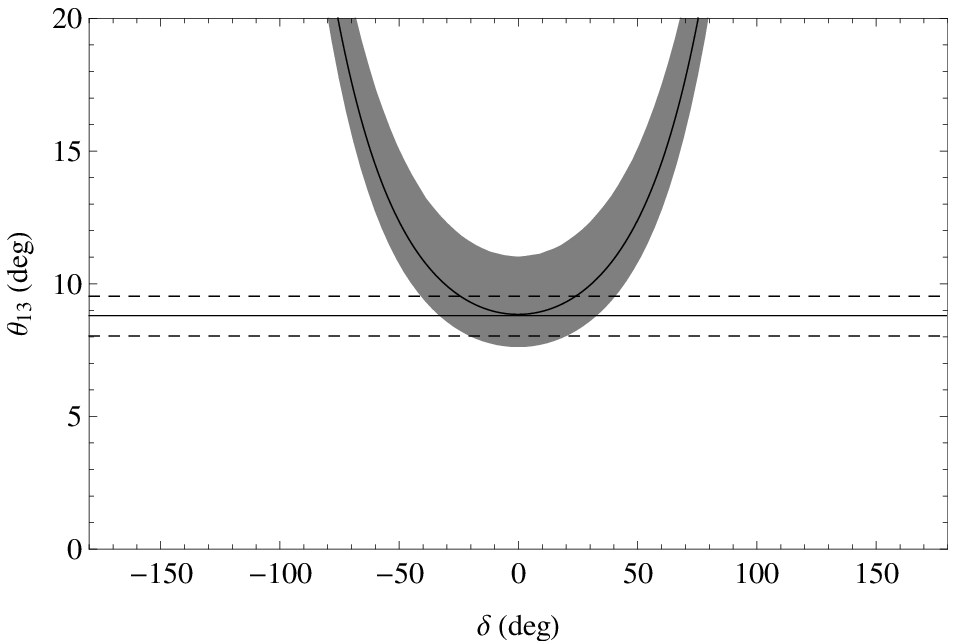}
        \end{subfigure}
        \caption{The $2\sigma$ allowed regions (shaded bands) in the ($\delta$, $\theta_{13}$), ($\delta$, $\theta_{12}$), and \mbox{($\delta$, $\theta_{23}$)} planes using measurements (with uncertainties) of the other two neutrino mixing angles and the CKM parameters. The solid curves within the shaded bands are the model predictions for the best-fit values of the other two mixing angles and the CKM parameters. The horizontal solid lines mark the best-fit values and the horizontal dashed lines mark the $2\sigma$ limits of $\theta_{23}$, $\theta_{12}$ and $\theta_{13}$.}
\label{fg:delta}
\end{figure}

\end {document}